\documentclass[pdftex,10pt,prd,aps,twocolumn,showpacs,floatfix]{revtex4-1}
\usepackage{graphicx}

\usepackage[english]{babel}
\usepackage{amsmath}
\usepackage{amssymb}

\usepackage[pdftitle={Formation of Stars},
pdfauthor={P. G. Miedema}, pdfsubject={Gauge
Invariant Cosmological Density Perturbations, Star Formation},
pdfstartview={FitH},
pdfkeywords={Star Formation, Cosmic Background Fluctuations},
bookmarks, bookmarksnumbered, colorlinks, linkcolor={blue},
citecolor={blue}]{hyperref}

\renewcommand{\vec}[1]{\mbox{\boldmath$#1$}}

\newcommand{\nul}{{\mbox{$\scriptscriptstyle(0)$}}}
\newcommand{\een}{{\mbox{$\scriptscriptstyle(1)$}}}

\newcommand{\gi}{{\scriptstyle\text{gi}}}
\newcommand{\dif}{\text{d}}

\newcommand{\mi}{\text{i}}

\begin{document}


\title{Formation of Population III Stars in a flat FLRW Universe}


\author{P.\ G.\ Miedema}
\email{pieter.miedema@gmail.com}
\affiliation{Netherlands Defense Academy\\
   Hogeschoollaan 2, \\ NL-4818 CR  Breda, The Netherlands}

\date{September 20, 2010}

\begin{abstract}
  Contrarily to general believe, a first-order cosmological
  perturbation theory based on Einstein's General Theory of Relativity
  explains the formation of massive primeval stars in a flat
  Friedmann-Lema\^\i tre-Robertson-Walker universe after decoupling of
  matter and radiation, whether or not Cold Dark Matter is present.
  The growth rate of a density perturbation depends on the heat loss
  of a perturbation during the contraction, but is independent of the
  particle mass.  The relativistic Jeans mass does depend on the
  particle mass.  If the Cold Dark Matter particle mass is equal to
  the proton mass, then the relativistic Jeans mass is equal to
  3500~solar masses, whereas the classical Jeans mass is a factor~145
  larger.
\end{abstract}


\maketitle


\section{Introduction}

A manifestly covariant gauge-invariant cosmological perturbation
theory for \textsc{flrw} universes based on the Theory of General
Relativity combined with Thermodynamics and a realistic equation of
state for the pressure $p=p(n,\varepsilon)$ has been developed in a
foregoing article~\cite{2010arXiv1003.4531M}.  In fact, in this
article the pioneering work of Lifshitz and
Khalatnikov~\cite{lifshitz1946,c15} has been redone, but now based on
two newly introduced gauge invariant quantities. These quantities,
which we baptized $\varepsilon_\een^\gi$ and $n^\gi_\een$, turned out
to be the energy density and the particle number density
perturbations.  Indeed, taking the non-relativistic limit
$v/c\rightarrow0$ the complete set of relativistic perturbation
equations reduce to the complete set of Newtonian equations
\begin{equation}
  \label{eq:newton}
  \nabla^2\varphi(\vec{x})=4\pi G\dfrac{\varepsilon^\gi_\een(\vec{x})}{c^2},
  \quad \varepsilon^\gi_\een(\vec{x})=n^\gi_\een(\vec{x})mc^2,
\end{equation}
thus identifying $\varepsilon_\een^\gi$ and $n^\gi_\een$ as the real
energy density and particle number density perturbations.

In the present article it will be demonstrated that the formation of
primordial stars, the so-called (hypothetical) population~\textsc{iii}
stars~\cite{2004ARA&A..42...79B,glover-2008},
can be understood with the revised Lifshitz-Khalatnikov perturbation
theory, whether or not Cold Dark Matter (\textsc{cdm}) is present.

\subsection{Former Results: Standard Perturbation Theory}
\label{sec:former-results}

It is generally accepted that in a universe filled with only `ordinary
matter,' i.e., elementary particles and photons but no \textsc{cdm},
linear perturbation theory predicts a too small growth rate to explain
the formation of structure in the universe.  The reason put forward in
the literature on structure formation is that in the linear phase of
the growth of an \emph{adiabatic} relative density perturbation
$\delta(t,\vec{x})$ in the era after decoupling of radiation and
matter, given by
\begin{equation}\label{eq:standard-growth}
\delta(t)=\delta(t_{\text{dec}})
\left(\dfrac{t}{t_{\text{dec}}}\right)^{2/3}, \quad t_{\text{dec}}
\le t \le t_{\text{p}},
\end{equation}
is insufficient for relative density perturbations as small as the
observed initial value $\delta(t_\text{dec})\approx10^{-5}$ to reach
the non-linear phase for times $t\le t_\text{p}$, where
$t_{\text{p}}=13.75\,\text{Gyr}$, the present age of the universe, and
$t_{\text{dec}}=381\,\text{kyr}$, the time of decoupling of matter and
radiation~\cite{2010arXiv1001.4744J,2010arXiv1001.4758B}.  This
generally accepted conclusion follows from the standard evolution
equation for density perturbations in a universe which is after
decoupling of matter and radiation assumed to be filled with a perfect
and pressure-less fluid (usually referred to as `dust') with equations
of state for the energy density $\varepsilon$ and pressure $p$
\begin{equation}
  \label{eq:newton-p0}
  \varepsilon=nmc^2, \quad p=0,
\end{equation}
where $n$ is the particle number density and $m$ the particle mass.

Before decoupling, Thomson scattering between photons and electrons
and Coulomb interactions between electrons and baryons were so rapid
that the photons and baryons are tightly coupled so that the
photon-baryon system behaves as a single fluid.  The standard
perturbation theory predicts that density perturbations in this
baryon-photon fluid oscillates with a \emph{constant} amplitude and
thus do not grow at all before decoupling.  Since \textsc{cdm} is
supposed to be electrically neutral, it is not linked by Coulomb
interactions to the baryon-photon fluid.  Therefore,
researchers~\cite{frenk-2002-300,springel-2006-440} in the field of
structure formation have assumed in their simulations that
\textsc{cdm} would have already clustered before decoupling and thus
would have formed seeds for baryon contraction after decoupling.  If
this would be true, then the slow growth (\ref{eq:standard-growth})
would be sufficient to explain structure in the universe.  Thus, the
mechanism of structure formation relies heavily on the particular
property of \textsc{cdm} before decoupling, namely that density
perturbations in \textsc{cdm} are not \emph{electrically} coupled to
perturbations in the total energy density.

\subsection{Results from the Revised Lifshitz-Khalatnikov Perturbation Theory}
\label{sec:new-pert-results}

However, in a foregoing article~\cite{2010arXiv1003.4531M} it has been
demonstrated that in the radiation-dominated era density perturbations
in ordinary matter and \textsc{cdm} are both \emph{gravitationally}
coupled to density perturbations in the total energy density.
Moreover, it has been found that small-scale density perturbations
oscillate in the radiation-dominated era with an \emph{increasing}
amplitude, proportional to $t^{1/2}$.  These energy (i.e., radiation,
ordinary matter and \textsc{cdm} tightly coupled together) density
perturbations will form the seeds for star formation after decoupling.
They manifest themselves as small temperature fluctuations
(\ref{eq:T-gamma}) in the cosmic background radiation.

Finally, it has been shown that neglecting the kinetic energy density
$\tfrac{3}{2}nk_{\text{B}}T$ with respect to the rest energy density
$nmc^2$ yields in the \emph{perturbed} universe the non-relativistic
limit with $\varepsilon^\gi_\een=n^\gi_\een mc^2$ and
$p_\nul=p^\gi_\een=0$, (\ref{eq:newton-p0}), implying that
$\delta_\varepsilon=\delta_n$.  Perturbations described by equations
of state (\ref{eq:newton-p0}) are adiabatic.  The growth is in this
case given by the standard expression (\ref{eq:standard-growth}),
which is far too slow to account for structure in the universe.  Since
adiabatic density perturbations cannot lose their internal energy to
their environment, they grow only under the influence of gravitation.
This explains their slow growth.

In the \emph{dark ages} of the universe (i.e., the epoch between
decoupling and the ignition of the first stars) a density perturbation
from which stars will eventually be formed should have initially a
somewhat smaller internal pressure than its environment and has to
lose some of its heat energy in order to grow faster than given by the
standard growth rate (\ref{eq:standard-growth}).  It has been
established~\cite{2010arXiv1003.4531M} that in a \emph{non}-static
universe density perturbations described by an equation of state
$p=p(n,\varepsilon)$ are \emph{diabatic}, whereas perturbations
described by (\ref{eq:newton-p0}) are adiabatic.  Therefore, the
evolution of density perturbations should be studied by using a
realistic equation of state for the pressure of the form
$p=p(n,\varepsilon)$, so that next to gravitational forces, also the
heat exchange of a perturbation can be incorporated into the
perturbation theory.  In fact, incorporating the realistic equations
of state (\ref{state-mat}), yields in the final dynamical perturbation
equation (\ref{dde-dn-de}) a source term and an evolution equation
(\ref{eq:dn-de}) for this source term.  Using the combined First and
Second Laws of Thermodynamics $\dif E=T\dif S-p\dif V+\mu\dif N$ the
source term of equation (\ref{dde-dn-de}) can be identified with the
entropy of a perturbation.  From equation (\ref{eq:dust-dimless}) one
may infer that the pressure term and the entropy term are of the same
order of magnitude.  This yields in the early stages of the
contraction of a perturbation a somewhat larger growth rate than in
the adiabatic case (\ref{eq:standard-growth}). This faster growth is
just enough for density perturbations with initial values as small as
$\delta_n\approx\delta_\varepsilon\lesssim10^{-5}$ to reach the
non-linear regime within $10^2$--$10^3\,\text{Myr}$.

\section{Outline}
\label{sec:outline}

Since \textsc{cdm} and ordinary matter particles behave
\emph{gravitationally} in exactly the same way and since the mass of a
\textsc{cdm} particle is as yet unknown, we assume that the
\textsc{cdm} particle mass is approximately equal to the proton mass.

After decoupling the cosmic fluid is a mixture of baryons (protons)
and \textsc{cdm}.  This mixture can be considered as a
non-relativistic monatomic perfect gas with equations of state for the
energy density and the pressure
\begin{equation}
  \varepsilon(n,T) = nmc^2+\tfrac{3}{2}nk_\mathrm{B}T, \quad
  p(n,T) = nk_{\mathrm{B}}T,    \label{state-mat}
\end{equation}
where $k_\mathrm{B}$ is Boltzmann's constant, $m$ the mean particle
mass, and $T$ the temperature of the matter.  It is assumed that
$m=m_{\text{H}}=m_{\text{CDM}}$, where $m_{\text{H}}$ is the
proton mass and $m_{\text{CDM}}$ the mass of a \textsc{cdm}
particle, implying that $mc^2\gg k_{\text{B}}T$ throughout the
matter-dominated era after decoupling.  Therefore, one may neglect the
pressure $nk_{\text{B}}T$ and kinetic energy density
$\tfrac{3}{2}nk_{\text{B}}T$ with respect to the rest-mass energy
density $nmc^2$ in the \emph{un}perturbed universe: the kinetic energy
of the particles in the universe has a negligible influence on the
global evolution of the universe.  Thus, as is well-known, the global
properties of the universe after decoupling are very well described by
a perfect and pressure-less fluid (`dust'), described by equations of
state (\ref{eq:newton-p0}).

At the moment of decoupling of matter and radiation photons could not
ionize matter any more and the two constituents fell out of thermal
equilibrium.  As a consequence, the pressure drops from a very high
radiation pressure $p=\tfrac{1}{3}a_{\mathrm{B}}T^4_\gamma$ just
before decoupling to a very low gas pressure $p=nk_{\mathrm{B}}T$
after decoupling.  This fast and chaotic transition from a high
pressure epoch to a very low pressure era may result locally in large
relative pressure perturbations.  These pressure perturbations will be
taken into account by incorporating the equations of state
(\ref{state-mat}) and their perturbed counterparts
\begin{equation}
  \label{eq:dn-de-dT}
  \delta_n-\delta_\varepsilon\approx
        -\dfrac{3}{2}\dfrac{k_{\text{B}}T_\nul}
        {mc^2}\delta_T, \quad
   \delta_p=\delta_n+\delta_T,
\end{equation}
into the new perturbation theory.  In the expressions
(\ref{eq:dn-de-dT}), $T_\nul$ is the background matter temperature and
$\delta_n$, $\delta_\varepsilon$, $\delta_T$ and $\delta_p$ are the
relative perturbations in the particle number density, the energy
density, the matter temperature and the pressure, respectively.  The
influence of pressure perturbations on the growth of small density
perturbations can only be investigated by using the revised
Lifshitz-Khalatnikov perturbation theory developed
in~\cite{2010arXiv1003.4531M}, since this theory not only has an
evolution equation (\ref{dde-dn-de}) for $\delta_\varepsilon$, but (in
contrast to all former perturbation theories) also an evolution
equation (\ref{eq:dn-de}) for the difference
$\delta_n-\delta_\varepsilon$.  This proves to be crucial for the
understanding of star formation in the early universe.  Although in a
linear perturbation theory $|\delta_p|\le1$ and $|\delta_T|\le1$, the
initial values of these quantities are, according to
(\ref{eq:dn-de-dT}), not constrained to be as small as the initial
values
\begin{equation}
  \label{eq:fluct-en}
\delta_\varepsilon(t_{\text{dec}},\vec{q})
\approx\delta_n(t_{\text{dec}},\vec{q})\lesssim10^{-5},
\end{equation}
as is demanded by
\textsc{wmap}-observations~\cite{2010arXiv1001.4744J,2010arXiv1001.4758B}. Since
the gas pressure $p=nk_{\text{B}}T$ is very low, its relative
perturbation $\delta_p\equiv p^\gi_\een/p_\nul$ and, accordingly, the
matter temperature perturbation $\delta_T\equiv T^\gi_\een/T_\nul$
could be large.

\section{Results}
\label{sec:results}

Just after decoupling, ordinary matter is mixed with \textsc{cdm}.  It
is found that the growth rate is independent of the particle mass,
i.e., the gravitational mechanism for star formation works equally
well with or without \textsc{cdm}.

It will be shown that just after decoupling at $z=1091$ negative
relative matter temperature perturbations as small as $-0.5\%$ yields
massive stars within $13.75\,\text{Gyr}$. The very first stars, the
so-called Population \textsc{iii} stars, come into existence between
$10^2\,\text{Myr}$ and $10^3\,\text{Myr}$. The star masses are in the
range from $4\times10^2\,\text{M}_\odot$ to $10^5\,\text{M}_\odot$,
with a peak around $3.5\times10^3\,\text{M}_\odot$. Density
perturbations with masses smaller than $3.5\times10^3\,\text{M}_\odot$
become non-linear at later times, because their internal gravity is
weaker. On the other hand, density perturbations with masses larger
than $3.5\times10^3\,\text{M}_\odot$ enter the non-linear regime also
later, since they do not cool down so fast due to their large
scale. The mass $3.5\times10^3\,\text{M}_\odot$ corresponds initially
to a scale of $6.2\,\text{pc}$.  These conclusions are outlined in
Figure~\ref{fig:collapse}. From this figure it follows that the growth
rate rapidly decreases for perturbations with masses below
$3.5\times10^3\,\text{M}_\odot$.  Therefore, the peak values in
Figure~\ref{fig:collapse} can be considered as the relativistic
counterparts of the classical \emph{Jeans mass}. However, the Jeans
mass does depend on the particle mass: heavier particles yield lighter
primordial stars.

\section{Basic Equations}
\label{sec:hierarchical}

For the equations of state (\ref{state-mat}) the background equations
for a flat ($k=0$) \textsc{flrw} universe with a vanishing
cosmological constant ($\Lambda=0$) reduce to
\begin{equation}
\label{subeq:einstein-flrw}
  3H^2 =\kappa\varepsilon_\nul, \quad
        \dot{\varepsilon}_\nul = -3H\varepsilon_\nul, \quad
      \dot{n}_\nul = -3Hn_\nul,
\end{equation}
where it is used that $w\equiv p_\nul/\varepsilon_\nul\ll1$, so that
the background pressure $p_\nul$ can be neglected with respect to the
background energy density $\varepsilon_\nul$. An overdot denotes
differentiation with respect to $ct$, and $\kappa\equiv8\pi G/c^4$.

It has been shown in a foregoing article~\cite{2010arXiv1003.4531M}
that for equations of state (\ref{state-mat}) and their perturbed
counterparts (\ref{eq:dn-de-dT}) the perturbation equations reduce to
\begin{subequations}
  \label{final-dust}
  \begin{align}
   & \ddot{\delta}_\varepsilon + 3H\dot{\delta}_\varepsilon-
  \left[\beta^2\frac{\nabla^2}{a^2}+
   \tfrac{5}{6}\kappa\varepsilon_\nul\right]
   \delta_\varepsilon=-\frac{2}{3}\frac{\nabla^2}{a^2}(\delta_n-\delta_\varepsilon),
   \label{dde-dn-de}\\
   & \frac{1}{c}\frac{\dif}{\dif t}
   \left(\delta_n-\delta_\varepsilon\right)=
   -2H\left(\delta_n-\delta_\varepsilon\right). \label{eq:dn-de}
  \end{align}
\end{subequations}
The quantity $\beta(t)$ defined by
$\beta\equiv\sqrt{\dot{p}_\nul/\dot{\varepsilon}_\nul}$ is, to a good
approximation, given by
\begin{equation}
     \beta(t) \approx \frac{v_\mathrm{s}(t)}{c}=\sqrt{\frac{5}{3}
        \frac{k_\mathrm{B}T_\nul(t)}{mc^2}}, \quad
      T_\nul\propto a^{-2},
\label{coef-nu1}
\end{equation}
with $v_\mathrm{s}$ the adiabatic speed of sound and $T_\nul$ the matter temperature.
Equation (\ref{eq:dn-de}) implies
\begin{equation}
  \label{eq:dn-dn-a-2}
  \delta_n-\delta_\varepsilon \propto a^{-2},
\end{equation}
where it is used that $H\equiv\dot{a}/a$, with $a(t)$ the scale factor
of the universe.  Combining (\ref{coef-nu1}) and (\ref{eq:dn-dn-a-2})
one gets from~(\ref{eq:dn-de-dT})
\begin{equation}
  \label{eq:dT-constant}
  \delta_T(t,\vec{x})\approx\delta_T(t_0,\vec{x}),
\end{equation}
to a very good approximation.  Using the well-known solutions of the
background equations (\ref{subeq:einstein-flrw})
\begin{equation}
  \label{eq:exact-sol-mat}
   H\propto t^{-1}, \quad  \varepsilon_\nul\propto t^{-2}, \quad
   n_\nul\propto t^{-2}, \quad a\propto t^{2/3},
\end{equation}
and substituting $\delta(t,\vec{x})=\delta(t,\vec{q})\exp(\mi\vec{q}\cdot\vec{x})$,
equations (\ref{final-dust}) can be combined into one equation
\begin{equation}\label{eq:dust-dimless}
    \delta_\varepsilon^{\prime\prime}+\frac{2}{\tau}\delta_\varepsilon^\prime+
\left[\dfrac{4}{9}\dfrac{\mu_\mathrm{m}^2}{\tau^{8/3}}-\frac{10}{9\tau^2}
\right]\delta_\varepsilon=
-\dfrac{4}{15}\dfrac{\mu^2_\mathrm{m}}{\tau^{8/3}}
\delta_T(t_0,\vec{q}),
\end{equation}
where $\tau\equiv t/t_0$ and a prime denotes differentiation with
respect to~$\tau$. The parameter $\mu_{\text{m}}$ is given~by
\begin{equation}\label{eq:const-mu}
\mu_\mathrm{m}\equiv\frac{2\pi}{\lambda_0}\frac{1}{H(t_0)}\frac{
v_\mathrm{s}(t_0)}{c},  \quad \lambda_0\equiv\lambda a(t_0),
\end{equation}
where $\lambda_0\equiv2\pi/|\vec{q}_0|$ is the scale of a perturbation
at time~$t_0$.  The constant $\mu_{\text{m}}$ can be expressed in
observable quantities.  To that end we use that the redshift $z(t)$ as
a function of the scale factor $a(t)$ is given by
\begin{equation}
  \label{eq:redshift}
  z(t)=\dfrac{a(t_{\text{p}})}{a(t)}-1,
\end{equation}
where $a(t_{\text{p}})$ is the present value of the scale factor. For
a flat \textsc{flrw} universe one may take $a(t_{\text{p}})=1$. Using
(\ref{eq:exact-sol-mat}), (\ref{eq:redshift}) and $T_\nul\propto
a^{-2}$, we get
\begin{equation}
  \label{eq:H-dec-wmap}
  \mu_{\mathrm{m}}=\dfrac{2\pi}{\lambda_0}
     \dfrac{\sqrt{\dfrac{5}{3}
    \dfrac{k_{\text{B}}T_{\nul}(t_{\text{dec}})}{m}}}
    {\mathcal{H}(t_{\text{p}})\bigl[z(t_{\text{dec}})+1\bigr]\sqrt{z(t_0)+1}},
\end{equation}
where $t_0$ is the time when a perturbation starts to
contract.  This expression is invariant under the replacement
$m\rightarrow\alpha m$ and
$\lambda_0\rightarrow\lambda_0/\sqrt{\alpha}$, for some
constant~$\alpha>0$. This implies that a perturbation
$\delta_\varepsilon$ with initial scale $\lambda_0$ in a cosmic fluid
with mean particle mass~$m$, evolves in exactly the same way as a
perturbation with initial scale $\lambda_0/\sqrt{\alpha}$ in a fluid
with mean particle mass~$\alpha m$.  In other words, the growth rate
is independent of the particle mass.

The mass $M(t_0)$ of a spherical density perturbation with
radius $\tfrac{1}{2}\lambda_0$ is given by
\begin{equation}
  \label{eq:M-dec}
  M(t_0)=
     \dfrac{4\pi}{3}\left(\tfrac{1}{2}\lambda_0\right)^3
     n_\nul(t_0)m.
\end{equation}
The particle number density $n_\nul(t_0)$ can be calculated from its
value $n_\nul(t_{\text{eq}})$ at the end of the radiation-dominated
era.  By definition, at the end of the radiation-domination era the
matter energy density $n_\nul mc^2$ equals the energy density of the
radiation:
\begin{equation}\label{eq:mat-en-eq}
n_\nul(t_\mathrm{eq})mc^2=a_\mathrm{B}T_{\nul\gamma}^4(t_\mathrm{eq}).
\end{equation}
Since $n_\nul\propto a^{-3}$ and $T_{\nul\gamma}\propto a^{-1}$, we
find, using (\ref{eq:redshift}), the particle number density at
time~$t_0$
\begin{equation}
  \label{eq:n-nul-t-dec}
  n_\nul(t_0)=
   \dfrac{a_{\mathrm{B}}T_{\nul\gamma}^4(t_{\mathrm{p}})}{mc^2}
   \bigl[z(t_{\mathrm{eq}})+1\bigr]\bigl[z(t_0)+1\bigr]^3.
\end{equation}
Combining (\ref{eq:M-dec}) and (\ref{eq:n-nul-t-dec}), we get for the
mass of a spherical density perturbation
\begin{equation}
  \label{eq:M-dec-n-dec}
  M(t_0)=
  \dfrac{4\pi}{3}\left(\tfrac{1}{2}\lambda_0\right)^3
 \dfrac{a_{\mathrm{B}}T_{\nul\gamma}^4(t_{\mathrm{p}})}{c^2}
   \bigl[z(t_{\mathrm{eq}})+1\bigr]\bigl[z(t_0)+1\bigr]^3.
\end{equation}
With the help of this expression the initial scale~$\lambda_0$ of a
perturbation is related to its mass at the initial time~$t_0$.

The influence of the mean particle mass $m$ on the mass $M(t_0)$ of a
primordial star can be studied by replacing $m$ by $\alpha m$ and
$\lambda_0$ by $\lambda_0/\sqrt{\alpha}$ ($\alpha>0$) in expressions
(\ref{eq:H-dec-wmap})--(\ref{eq:n-nul-t-dec}).  It is found from
(\ref{eq:M-dec-n-dec})
\begin{equation}
  \label{eq:m-propto}
  M(t_0)\propto \alpha^{-3/2}.
\end{equation}
In other words, the heavier the particles in the universe, the lighter
the primordial stars.  For example, if $\alpha=10$
then the mean particle is $m=10m_{\text{H}}$, implying that the Jeans
mass in Figure~\ref{fig:collapse} becomes approximately
$10^2\,\text{M}_\odot$.

Finally, the influence of the initial time on a star mass is
determined.  It follows from (\ref{eq:H-dec-wmap}) that
\begin{equation}
  \label{eq:lambda-z0}
  \lambda_0 \propto \left[z(t_0)+1\right]^{-1/2},
\end{equation}
implying with (\ref{eq:M-dec-n-dec}) that
\begin{equation}
  \label{eq:MJ-z0}
  M(t_0) \propto \left[z(t_0)+1\right]^{3/2}.
\end{equation}
Thus, the later a perturbation starts to contract, the smaller the
mass will be.  For example, if a perturbation starts to contract at
$z(t_0)=1$, then the Jeans mass in Figure~\ref{fig:collapse} will be
$M_{\text{J}}(t_0)\approx0.27\,\text{M}_\odot$.

\section{Initial values from WMAP}
\label{sec:wmap}

The physical quantities measured by
\textsc{wmap}~\cite{2010arXiv1001.4744J,2010arXiv1001.4758B} and
needed in the present theory of primordial star formation are the
redshifts at matter-radiation equality and decoupling, the present
values of the Hubble function and the background radiation
temperature, the age of the universe and the fluctuations in the
background radiation temperature:
\begin{subequations}
\label{subeq:wmap}
\begin{align}
   z(t_{\text{eq}}) & = 3196, \label{eq:z-eq} \\
   z(t_{\text{dec}})& = 1091,  \label{eq:init-wmap-z} \\
   cH(t_{\text{p}})=\mathcal{H}(t_{\text{p}})& =
   71.0\,\text{km/sec/Mpc},\label{eq:init-wmap-T}\\
  T_{\nul\gamma}(t_\text{p})& = 2.725\,\text{K}, \label{eq:back-T-tp}\\
  t_{\text{p}} & =13.75\text{ Gyr}, \\
  \delta_{T_\gamma}(t_{\text{dec}}) & \lesssim 10^{-5}. \label{eq:T-gamma}
\end{align}
\end{subequations}
At decoupling the matter temperature is equal to the radiation
temperature.  The latter can be calculated from the fact that
$T_{\nul\gamma}\propto a^{-1}$ and the quantities
(\ref{eq:init-wmap-z}) and (\ref{eq:back-T-tp}). Using
(\ref{eq:redshift}), one finds for the matter temperature at
decoupling
\begin{equation}
\label{eq:decoup-temp}
    T_\nul(t_{\text{dec}})= T_{\nul\gamma}(t_{\text{dec}})=2976\,\text{K}.
\end{equation}
Substituting the observed values (\ref{eq:init-wmap-z}) and
(\ref{eq:init-wmap-T}) into (\ref{eq:H-dec-wmap}), one gets, using
also (\ref{eq:decoup-temp}),
\begin{equation}\label{eq:nu-m-lambda}
    \mu_\mathrm{m}=\dfrac{518.5}{\lambda_0\sqrt{z(t_0)+1}}, \quad
\lambda_0 \text{ in pc},
\end{equation}
where we have used that
$1\,\mathrm{pc}=3.0857\times10^{16}\,\mathrm{m}$
($1\,\mathrm{pc}=3.2616\;\mathrm{ly}$).

Finally, using that one solar mass is
$1.98892\times10^{30}\,\text{kg}$, we find from (\ref{subeq:wmap})
that
\begin{equation}
  \label{eq:M-dec-solar}
  M(t_0)=1.148\times10^{-8}\lambda^3_0
      \bigl[z(t_0)+1\bigr]^3\,\text{M}_\odot.
\end{equation}
The expression (\ref{eq:M-dec-solar}) will be used to convert the
scale $\lambda_0$ (expressed in units of $1\,\text{pc}$) of a perturbation, which starts to contract at a
redshift of $z(t_0)$, into its mass $M(t_0)$ (expressed in units of
the solar mass).

\section{Population III Star Formation}
\label{sec:hier-struc}

In this section the evolution equation (\ref{eq:dust-dimless}) is
solved numerically. To that end the differential equation solver
\texttt{lsodar} with root finding capabilities is used. This solver is
included in the package \texttt{deSolve}~\cite{soetaert-2010}, which,
in turn, is included in~\textsf{R}, a system for statistical
computation and graphics~\cite{R}. Star formation which starts at
cosmological redshift $z=1091$, i.e., at $t_0=t_{\text{dec}}$, is
investigated.

\begin{figure}[h]
\begin{center}
\includegraphics[scale=0.48]{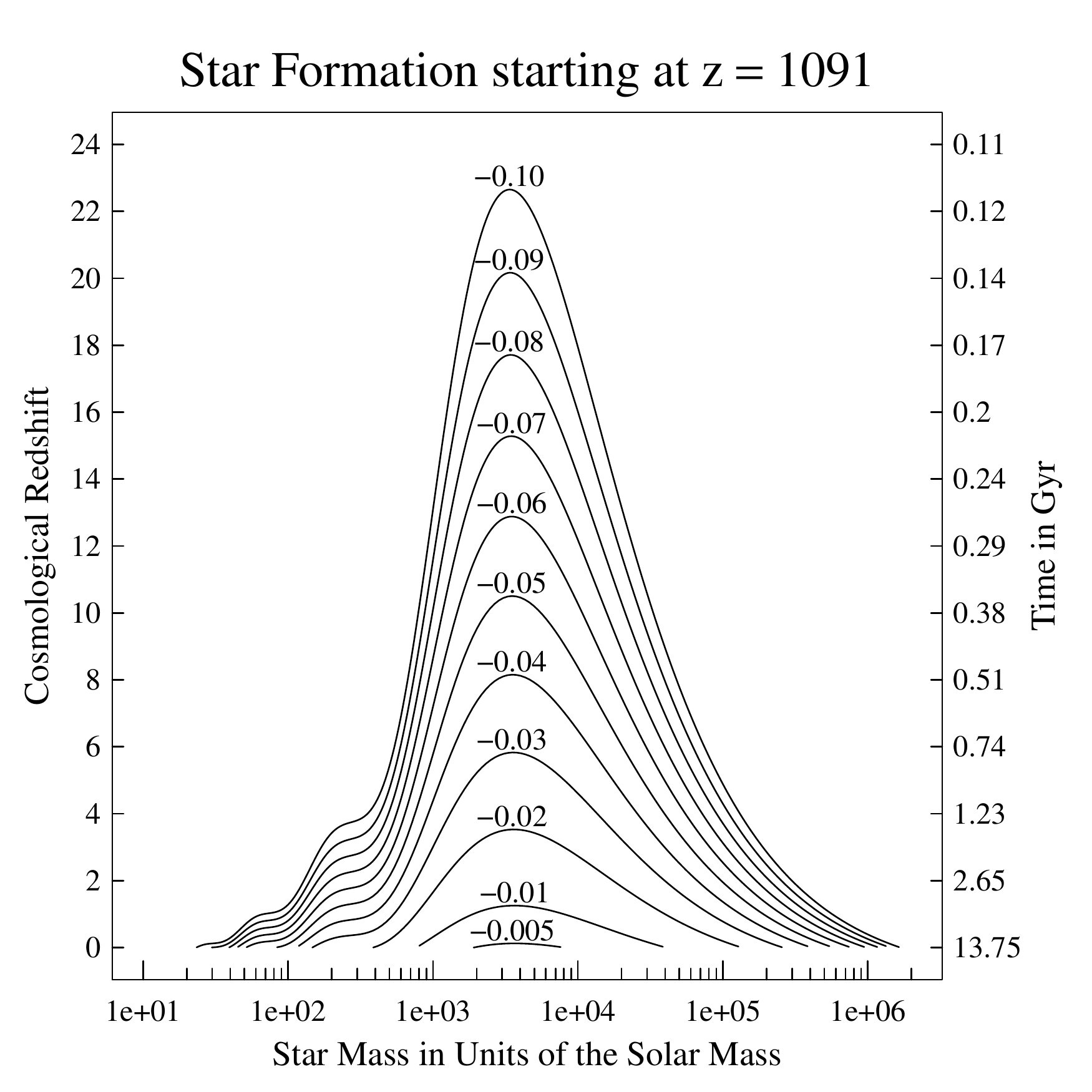}
\caption{The curves give the redshift at which a linear perturbation
  in the particle number density with initial values
  $\delta_n(t_{\text{dec}},\vec{q})\approx\delta_\varepsilon(t_{\text{dec}},\vec{q})\approx10^{-5}$
  and $\delta^{\prime}_n(t_{\text{dec}},\vec{q})=0$ starting to grow
  at an initial redshift of $z(t_{\mathrm{dec}})=1091$ becomes
  non-linear, i.e., $\delta_n\approx\delta_\varepsilon\approx1$. During the evolution we have
  $\delta_p(t,\vec{q})=\delta_T(t_{\text{dec}},\vec{q})+\delta_n(t,\vec{q})$.
  The numbers at each of the curves are the initial relative
  perturbations in the matter temperature
  $\delta_T(t_{\mathrm{dec}},\vec{q})$. For each curve, the Jeans mass
  (i.e., the peak value) is at $3.5\times10^3\,\mathrm{M}_\odot$.}
\label{fig:collapse}
\end{center}
\end{figure}

The \textsc{wmap} observations of the fluctuations in the background
radiation temperature yield for the fluctuations in the energy density
and particle number density (\ref{eq:fluct-en}).  In addition, it is
assumed that
\begin{equation}
  \label{eq:d-fluct-en-dt}
  \dot{\delta}_\varepsilon(t_{\text{dec}},\vec{q}) \approx 0,
\end{equation}
i.e., during the transition from the 
radiation-dominated era to the era after decoupling energy density
perturbations are approximately constant with respect to time.

Figure~\ref{fig:collapse} has been constructed as follows.  For each
choice of $\delta_T(t_{\text{dec}},\vec{q})$ equation
(\ref{eq:dust-dimless}) is integrated for a large number of values for
the initial scales $\lambda_0=\lambda_{\text{dec}}$, using the initial
values (\ref{eq:fluct-en}) and (\ref{eq:d-fluct-en-dt}).  The
integration starts at $\tau\equiv t/t_{\text{dec}}=1$, i.e., at
$z=z(t_{\text{dec}})$ and will be halted if either $z=0$ (i.e.,
$\tau=[z(t_{\text{dec}})+1]^{3/2}$), or
$\delta_\varepsilon(t,\vec{q})=1$ has been reached.  One integration
run yields one point on the curve for a particular choice of the scale
$\lambda_{\text{dec}}$ if $\delta_\varepsilon(t,\vec{q})=1$ has been
reached for $z>0$.  If the integration halts at $z=0$ and still
$\delta_\varepsilon(t,\vec{q})<1$, then the perturbation belonging to
that particular scale $\lambda_{\text{dec}}$ has not yet reached its
non-linear phase today, i.e., at $t_{\text{p}}=13.75\,\text{Gyr}$. On
the other hand, if the integration is stopped at
$\delta_\varepsilon(t,\vec{q})=1$ and $z>0$, then the perturbation has
become non-linear within $13.75\,\text{Gyr}$.

The above described procedure is repeated for
$\delta_T(t_{\text{dec}},\vec{q})$ in the range $-0.005, -0.01, -0.02,
\ldots, -0.1$. During the evolution, the relative pressure
perturbation evolves according to~(\ref{eq:dn-de-dT}) and
(\ref{eq:dT-constant}):
\begin{equation}
  \label{eq:evo-const-T}
  \delta_p(t,\vec{q})=\delta_T(t_{\text{dec}},\vec{q})+\delta_n(t,\vec{q}).
\end{equation}
The fastest growth is seen for perturbations with a mass of
approximately $3.5\times10^3\,\text{M}_\odot$.  This value is nearly
independent of the initial value of the matter temperature
perturbation $\delta_T(t_{\text{dec}},\vec{q})$. Even density
perturbations with an initial relative matter temperature perturbation
as small as $\delta_T(t_{\text{dec}},\vec{q})=-0.5\%$ reach their
non-linear phase at $z=0.13$ ($T_{\nul\gamma}=3.1\,\text{K}$,
$t=11.5\,\text{Gyr}$) provided that its mass is around
$3.5\times10^3\,\text{M}_\odot$.  Perturbations with masses smaller
than $3.5\times10^3\,\text{M}_\odot$ reach their non-linear phase at a
later time, because their internal gravity is weaker. On the other
hand, perturbations with masses larger than
$3.5\times10^3\,\text{M}_\odot$ cool down slower because of their
large scales, resulting also in a smaller growth rate.  Since the
growth rate decreases rapidly for perturbations with masses below
$3.5\times10^3\,\text{M}_\odot$, the latter mass will be considered as
the relativistic counterpart of the classical \emph{Jeans mass}. This
mass corresponds to a Jeans scale of
$6.2\,\text{pc}\approx20\,\text{ly}$.  This scale is much smaller than
the horizon size at decoupling, given by
$d_{\text{H}}(t_{\text{dec}})=3ct_{\text{dec}}\approx350\,\text{kpc}$.

\section{Heat Loss during Contraction}
\label{sec:heat-loss}

In this section the heat loss of a density perturbation during its
contraction is calculated. To that end the combined first second law
of thermodynamics (24) in Ref.~\cite{2010arXiv1003.4531M} is rewritten
in the form
\begin{equation}
  \label{eq:fs-therm-law}
  T_\nul
  s^\gi_\een=-\dfrac{\varepsilon_\nul}{n_\nul}(\delta_n-\delta_\varepsilon)-
  \dfrac{p_\nul}{n_\nul}\delta_n,
\end{equation}
where it is used that $w\equiv p_\nul/\varepsilon_\nul$.  Substituting
expressions (\ref{state-mat}) and (\ref{eq:dn-de-dT}) into
(\ref{eq:fs-therm-law}) and using also (\ref{eq:dT-constant}), one
finds the entropy per particle of a density perturbation:
\begin{equation}
  \label{eq:heat-exchange}
  s^\gi_\een(t,\vec{x}) \approx \tfrac{1}{2}k_{\text{B}}
  \bigl[3\delta_T(t_0,\vec{x})-2\delta_n(t,\vec{x})\bigr],
\end{equation}
where it is used that $mc^2\gg k_{\text{B}}T_\nul$.  For all values of
$\delta_T(t_0,\vec{x})$ in Figure~\ref{fig:collapse} and initial
values (\ref{eq:fluct-en}) the entropy perturbation is negative,
$s^\gi_\een<0$.  Since for growing perturbations one has
$\dot{\delta}_n>0$ the entropy perturbation decreases, i.e.,
$\dot{s}^\gi_\een=-k_{\text{B}}\dot{\delta}_n<0$, during contraction.
This implies that a growing perturbation loses a part of its internal
energy to its environment.  This is to be expected, since a local
density perturbation is not isolated from its environment.  Only for
an isolated system the entropy never decreases.

\section{Classical Jeans Mass}
\label{sec:class-JM}

In this section the classical Jeans mass, derived from the Newtonian
theory of gravity, is compared with the relativistic Jeans mass which
follows from the revised Lifshitz-Khalatnikov perturbation theory.

The classical Jeans length at time $t_0$ is given
by~\cite{weinberg-2008}
\begin{equation}
  \label{eq:class-Jeans-length}
  \lambda_{\text{J}}(t_0)=v_{\text{s}}(t_0)\sqrt{\dfrac{\pi}{Gn_\nul(t_0)m}}.
\end{equation}
Using (\ref{coef-nu1}), (\ref{eq:M-dec}) and (\ref{eq:n-nul-t-dec})
one finds for the classical Jeans mass at decoupling:
\begin{equation}
  \label{eq:class-Jeans-mass}
  M_{\text{J}}(t_{\text{dec}})\approx5.1\times10^5\,\text{M}_\odot,
\end{equation}
where the \textsc{wmap} values (\ref{subeq:wmap}) have been used.  The
classical Jeans mass (\ref{eq:class-Jeans-mass}) is much larger than
the relativistic Jeans mass, $3.5\times10^3\,\text{M}_\odot$ which
follows from the revised Lifshitz-Khalatnikov perturbation
theory. This difference is due to the fact that in the classical
perturbation theory based on the equations of state
(\ref{eq:newton-p0}) the heat loss of a perturbation is not taken into
account, whereas the effect of heat loss on the growth of a
perturbation is included in the revised Lifshitz-Khalatnikov
perturbation theory based on the equations of state (\ref{state-mat}).
In other words, since a perturbation loses some of its energy, gravity
can be somewhat weaker to make a perturbation contract.  The classical
Jeans mass (\ref{eq:class-Jeans-mass}) corresponds to a classical
Jeans scale of $32.4\,\text{pc}\approx106\,\text{ly}$. Just as is the
case for $\mu_{\text{m}}$ (\ref{eq:H-dec-wmap}), the expression
(\ref{eq:class-Jeans-length}) is invariant under the replacement
$m\rightarrow\alpha m$ and
$\lambda_{\text{J}}\rightarrow\lambda_{\text{J}}/\sqrt{\alpha}$, for
some constant~$\alpha>0$. As a consequence, the classical Jeans mass
$M_{\text{J}}$ is proportional to $\alpha^{-3/2}$, just as in the
relativistic case (\ref{eq:m-propto}).

Finally, the classical Jeans mass of a perturbation starting at
$z(t_0)=1$ follows from (\ref{eq:MJ-z0}) and
(\ref{eq:class-Jeans-mass}). It is found that
$M_{\text{J}}(t_0)\approx40\,\text{M}_\odot$.

\section{Conclusion}
\label{sec:conclusie}

Three important conclusions can now be drawn.  Firstly, there is no
need to make use of alternative gravitational theories: the Theory of
General Relativity explains the formation of massive primordial stars
in our universe.  In other words, Einstein's gravitational theory not
only describes the \emph{global} characteristics of the universe, but
is also \emph{locally} successful.  Secondly, although there is strong
evidence for the existence of \textsc{cdm}~\cite{CGM-2006}, it is not
needed for the formation of primeval stars.  Finally, is has
been demonstrated that not only for large-scale perturbations one
should use the theory of relativity, but also for small-scale
perturbations: because of the spurious gauge modes present in the
Newtonian theory of gravity, it fails to predict primordial stars.

\section*{Acknowledgments}
The author is indebted to Willem van Leeuwen for critically reading
the manuscript.


\bibliographystyle{unsrtnat}
\bibliography{/home/pieter/Documents/artikel/literatuur-database/diss}

\end{document}